\documentclass{elsarticle}

	\usepackage{bm}
	\usepackage{amsfonts}
	\usepackage{amssymb}
	\usepackage{amsmath}

 	\usepackage{graphicx}

	\usepackage{hyperref}		
	\usepackage[figure]{hypcap}		

	\hypersetup{
	   	bookmarks		= false,		
		unicode		= false,		
		pdfborder		= {0 0 0}		
		pdftoolbar		= true,		
		pdfmenubar		= true,		
		pdffitwindow		= false,     		
		pdfstartview		= {FitH},    		
		pdfnewwindow	= true,      		
		colorlinks		= true,		
		linkcolor		= blue,		
		citecolor		= blue,		
		filecolor		= blue,		
		urlcolor		= blue,		
		linktocpage		= true,		
		pdftitle		= {Geometry-dependent gapped states in suspended graphene flakes},
		pdfauthor		= {M. Mucha-Kruczynski, V.I. Fal'ko},
		pdfsubject		= {},
		pdfcreator		= {MikTeX},
		pdfproducer		= {Marcin Mucha-Kruczynski},
		pdfkeywords		= {inhomogeneous strain in graphene},
}

	\newcommand{\vect}[1]{\boldsymbol{#1}}

	\newcommand{\xmem}{x}
	\newcommand{\ymem}{y}
	\newcommand{\xcryst}{x'}
	\newcommand{\ycryst}{y'}
	\newcommand{\uxcryst}{{u}_{\xcryst}}
	\newcommand{\uycryst}{{u}_{\ycryst}}
	\newcommand{\uxmem}{{u}_{\xmem}}
	\newcommand{\uymem}{{u}_{\ymem}}
	\newcommand{\comments}[1]{}
	\newcommand{\pr}{\sigma}
	\newcommand{\rab}{r_{AB}}
	\newcommand{\parmx}{\partial_{\xmem}}
	\newcommand{\parmy}{\partial_{\ymem}}
	\newcommand{\parcx}{\partial_{\xcryst}}
	\newcommand{\parcy}{\partial_{\ycryst}}
	\newcommand{\parmxx}{\partial_{\xmem\xmem}}
	\newcommand{\parmyy}{\partial_{\ymem\ymem}}
	\newcommand{\parmxy}{\partial_{\xmem\ymem}}
	\newcommand{\B}{\mathcal{B}}

\begin{document}

\title{Pseudo-magnetic field distribution and pseudo-Landau levels in suspended graphene flakes}

\author[lancs]{M. Mucha-Kruczy\'{n}ski\corref{cor1}}
\ead{m.mucha-kruczynski@lancaster.ac.uk}

\author[lancs]{V.I. Fal'ko}

\cortext[cor1]{Corresponding author}
\address[lancs]{Department of Physics, Lancaster University,
Lancaster, LA1~4YB, UK}

\begin{abstract}
Combining the tight-binding approximation and linear elasticity theory for a planar membrane, we investigate stretching of a graphene flake assuming that two opposite edges of the sample are clamped by the contacts. We show that, depending on the aspect ratio of the flake and its orientation, gapped states may form in the membrane in the vicinity of the contacts. This gap in the pre-contact region should be biggest for the armchair orientation of the flake and $\frac{W}{L}\lesssim1$.
\end{abstract}
\maketitle

Over the past eight years since its successful isolation \cite{novoselov_science_2004}, graphene has attracted a lot of attention \cite{geim_science_2009, castro_neto_rmp_2009, abergel_advphys_2010}. In particular, fabrication of high quality graphene samples suspended over the substrate and thus decoupled from its influence \cite{du_natnano_2008, bolotin_ssc_2008} stimulated investigations of the closest vicinity of neutrality point in the electronic band structure looking for electron-electron interaction effects in both monolayer \cite{elias_natphys_2011} and bilayer \cite{feldman_natphys_2009, weitz_science_2010, martin_prl_2010, mayorov_science_2011} graphene. 

However, it is easy to imagine that such an atomically flat membrane placed over a ridge undergoes unintentional mechanical deformations during the fabrication process or operation of the device. At the same time, it is well-known that for graphene systems lattice distortions are coupled to electronic degrees of freedom resulting in gauge fields resembling valley-dependent effective magnetic field $\B$ acting on electrons in the deformed flake \cite{suzuura_physe_2000, suzuura_prb_2002, manes_prb_2007}. It has been shown that application of an electric field with a gate placed underneath the flake deforms it and leads to suppression of the conductance \cite{fogler_prl_2008}. Similarly, in a four-contact configuration the interaction between the side contacts and charge carriers in graphene distorts the flake and can impede observation of quantum Hall plateaus \cite{prada_prb_2010}.

In this paper, we combine the tight-binding approximation with the linear theory of elasticity and investigate distributions of the strain-induced pseudo-magnetic field with relation to the flake orientation and its aspect ratio. We show that for the zig-zag orientation of the flake, a $\B=0$ region connecting both contacts always exists in the middle of the flake. In contrast, for the armchair orientation, regions of nonzero pseudo-magnetic field are situated along the contacts and we highlight the possibility of strain-induced Landau level formation within those regions for samples with $W\lesssim L$.

\begin{figure}[b]
\centering
\includegraphics[width=0.5\columnwidth]{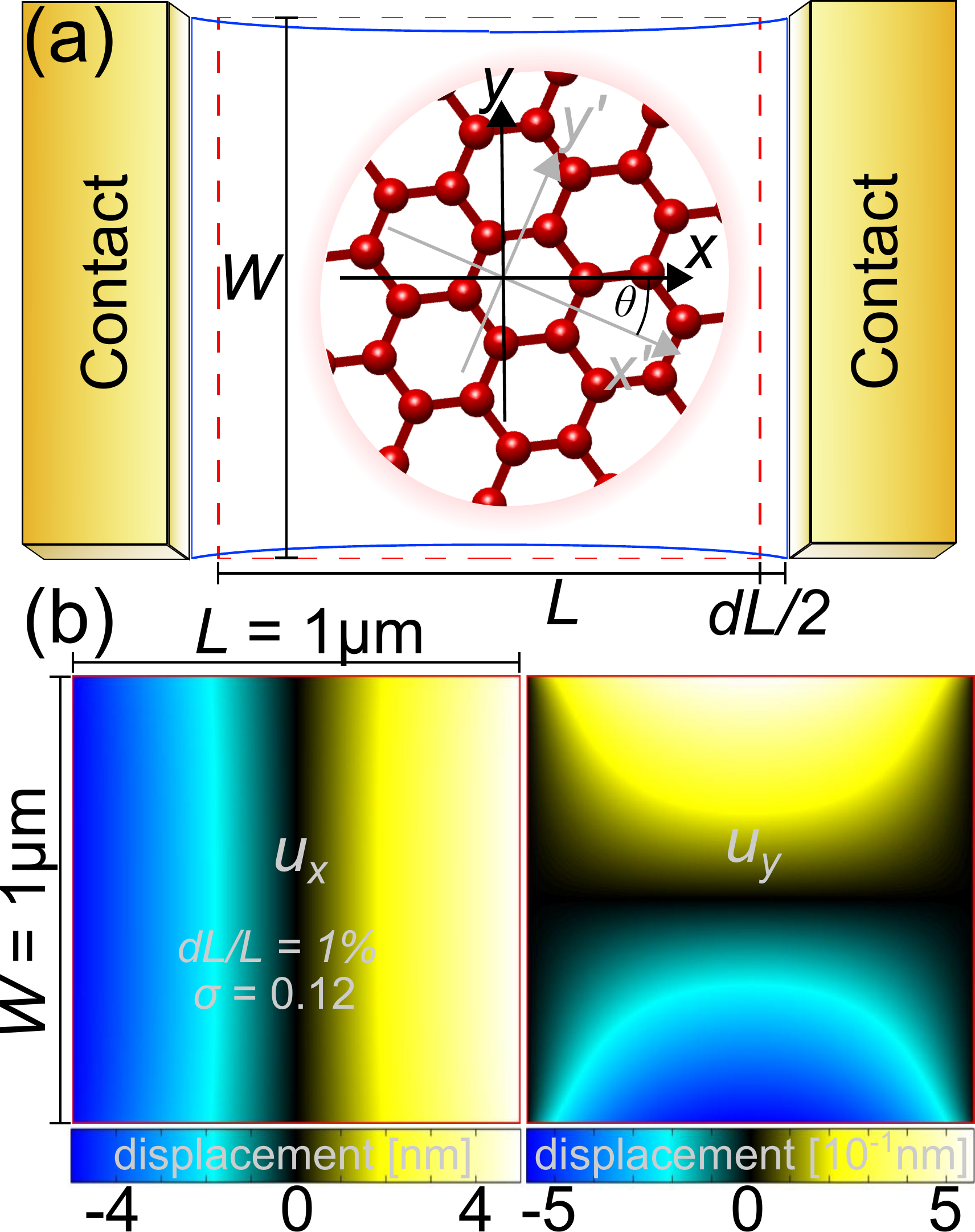}
\caption{(a) Rectangular graphene membrane under tension. The red dashed and blue solid line depict the membrane before and after the deformation, respectively. In the inset, the relation between the membrane and crystalline coordinates $xy$ and $x'y'$ is shown. (b) Distribution of the $\uxmem$ and $\uymem$ components of displacement for a square membrane.}
\label{fig:model}
\end{figure}

We model the suspended flake with a flat rectangular elastic membrane of length $L$ and width $W$ clamped at the edges by metallic contacts, as shown in Fig. \ref{fig:model}(a). We denote the membrane coordinate system as $\xmem\ymem$ and the crystalline coordinate system as $\xcryst\ycryst$. We choose the latter in such a way that for monolayer graphene the $\xcryst$ axis points along the armchair direction and call $\theta$ the angle between the $\xmem$ and $\xcryst$ axes.

After the deformation, the total elongation of the membrane along $\xmem$ is $dL$ ($\frac{dL}{2}$ on each side), leading to an inhomogeneous distribution of strain. Here, we are interested in the effects caused by small strains ($\lesssim 1\%$) and ignore any strain-induced wrinkling developing in the membrane. Then, the problem simplifies to that of stretching of a planar sheet. Within the linear elasticity theory \cite{landau_book_1970, timoshenko_book_1970}, the equations for displacement $\vect{u}(\xmem,\ymem)=(\uxmem,\uymem)$ of a point $(\xmem,\ymem)$ in the membrane take the form
\begin{align}\begin{split}\label{eqn:continuum_equations}
& 2\parmxx u_{\xmem} + (1-\pr)\parmyy u_{\xmem} + (1+\pr)\parmxy u_{\ymem} = 0, \\
& 2\parmyy u_{\ymem} + (1-\pr)\parmxx u_{\ymem} + (1+\pr)\parmxy u_{\xmem} = 0,
\end{split}\end{align}
with clamped boundary condition for the left and right edges and free boundary condition for the top and bottom edges,
\begin{align}\label{eqn:boundary_conditions}
\!\left\{ \begin{array}{l} \!\!\uxmem(\pm\frac{L}{2},\ymem)=\pm\frac{dL}{2} \\ \!\!\uymem(\pm\frac{L}{2},\ymem)=0 \end{array} \right.\!\!, \quad\!\! \left\{ \begin{array}{l} \!\![\pr\parmx\uxmem+\parmy\uymem]_{\ymem=\pm\frac{W}{2}} = 0 \\ \!\![\parmx\uymem+\parmy\uxmem]_{\ymem=\pm\frac{W}{2}} = 0 \end{array} \right. \!.
\end{align}
Above, $\pr$ is the Poisson's ratio for graphene and in this work, we take $\pr=0.12$ \cite{farjam_prb_2009, ribeiro_njp_2009, zakharchenko_prl_2009}.

However, microscopically the lattice is described in the crystalline coordinates $\xcryst\ycryst$. On this level and within the tight-binding picture, strain-induced changes in the distances between carbon atoms in the lattice lead to shifts of the energies of the electron $\pi$-orbitals, $\epsilon_{2p}$. This introduces the diagonal term $\frac{\epsilon_{2p}'\rab}{2}\nabla\cdot\vect{u}$, with $\epsilon_{2p}'=\frac{\partial\epsilon_{2p}}{\partial\rab}\approx 6$eV/\AA \cite{ferrone_ssc_2011} expressing the change of the on-site energy with growing bond length $\rab$ (for unperturbed lattice, $\rab=1.46\AA$). At the same time, change in the bond lengths introduces asymmetry in the couplings for the three nearest neighbours surrounding any given atomic site. The form of the resulting addition to the electronic Hamiltonian written for the states in the vicinity of one of the Brillouin zone corners (valleys) is equivalent to the addition of a valley-dependent gauge vector potential \cite{castro_neto_rmp_2009},
\begin{align}\label{eqn:gauge_potential}
\vect{\mathcal{A}}_{0}(\xcryst,\ycryst) = \xi\frac{\hbar\eta_{0}}{2e\rab}\left(\begin{array}{c} \parcx\uxcryst - \parcy\uycryst \\ -\parcx\uycryst-\parcy\uxcryst \end{array}\right), \,\,\,\B=\nabla\times\vect{\mathcal{A}}_{0}.
\end{align}
The parameter $\eta_{0}=\frac{\partial\ln\gamma_{0}}{\partial\ln\rab}$ expresses the change of the nearest neighbour coupling $\gamma_{0}$ with the change of the bond length $\rab$. Its value $\eta_{0}\approx -3$ can be estimated from the DFT calculations \cite{ferrone_ssc_2011} and electron-phonon interaction constant measured by Raman spectroscopy \cite{basko_prb_2009}. The $\uxcryst$ and $\uycryst$ are the components of the displacement in the crystalline coordinates, $\xi=\pm1$ identifies one of the two valleys and $e$ is electron charge. A strong enough inhomogeneous strain may cause quantization of electronic states into Landau levels \cite{guinea_natphys_2009, guinea_prb_2010, levy_science_2010} and introduce gaps into the electronic spectrum. With the help of the angle $\theta$ expressing the connection between the membrane and crystalline coordinate systems, $\B$ can be written in the macroscopic coordinates $\xmem\ymem$ as
\begin{align}\label{eqn:mag_field}
\B = \frac{\xi\hbar\eta_{0}}{2e\rab} \left\{ \sin\!3\theta\left[ \left(\parmxx-\parmyy\right)\uxmem\!-\!2\parmxy\uymem \right] \!-\! \cos\!3\theta\left[ 2\parmxy\uxmem\!+\!\left(\parmxx-\parmyy\right)\uymem \right] \right\}.
\end{align}

The problem of a planar membrane stretched like in Fig. \ref{fig:model}(a) does not have a simple analytical solution and hence we applied the finite element method \cite{zienkiewicz_book_2000} to obtain displacements in membrane coordinates. Due to the need of the first and second derivatives of displacement, a nine-point rectangular element has been used. An example of the $\uxmem$ and $\uymem$ components of the displacement for a membrane with length $L=1\mathrm{\mu}$m, aspect ratio $\frac{W}{L}=1$ and longitudinal strain $\frac{dL}{L}=1\%$ is presented in Fig. \ref{fig:model}(b)\footnote{A mesh of $400\times 400$ elements used in this example is a good representative of the mesh sizes used in this work.}. The first derivatives of the displacement which appear in the expression for the gauge potential $\vect{\mathcal{A}}_{0}$, Eq. \eqref{eqn:gauge_potential}, are associated with the components of the stress tensor. For planar problems treated within the linear theory of elasticity, stress singularities occur in sharp corners \cite{williams_jam_1952} due to the incompatibility of the boundary conditions at the corner point and nonlinear theory is required to obtain physical results in their vicinities \cite{sinclair_amr_2004_I, sinclair_amr_2004_II}. Hence, within our model, both $\vect{\mathcal{A}}_{0}$ and $\B$ diverge at the corners of the flake. These divergencies are only important within several lattice constants away from the corner \cite{vafek_prb_1999} and we ignore those regions in our considerations. Physically, the boundary conditions can be locally relaxed as the stress is limited by the treshold for sliding of the flake.

\begin{figure}[t]
\centering
\includegraphics[width=0.5\columnwidth]{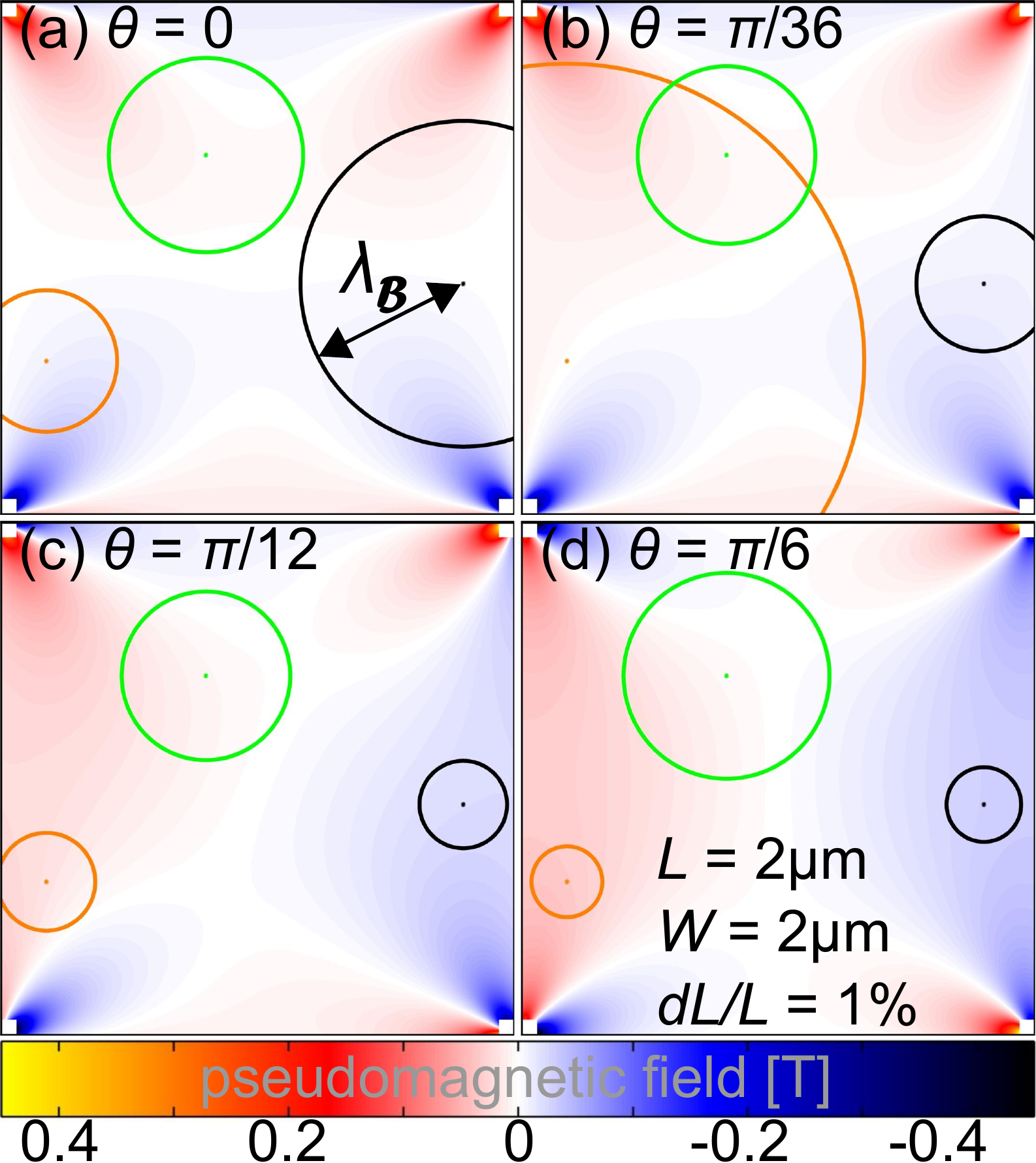}
\caption{Distribution of the pseudo-magnetic field $\B$ for electrons around the $K_{+}$ valley as a function of the angle $\theta$ for a square membrane with $L=2\mathrm{\mu}$m and $\frac{dL}{L}=1\%$. Note that the distribution of $\B$ in the vicinity of the sharp corners of the flake is not shown and replaced with white squares. The black, green and orange circles show the magnetic length for three points in the membrane marked by a dot in the same colour.}
\label{fig:pseudomag_field_theta}
\end{figure}

Having obtained the first derivatives of the displacement, one can compute the strain-induced on-site scalar potential $\Phi(x,y)=\frac{\epsilon_{2p}'\rab}{2}\nabla\cdot\vect{u}(x,y)$. However, as opposed to the vector potential $\vect{\mathcal{A}}_{0}$, the scalar potential is screened by carriers in the flake. Detailed investigations of this screening indicate that linear response theory approximates its effects reasonably well \cite{gibertini_prb_2010}. Within this scheme, the induced carrier density $\delta\!n(x,y)$, is \cite{gibertini_prb_2010}
\begin{align}
\delta\!n(x,y) = \mathcal{F}\left[ \frac{\chi(q)}{\epsilon(q)}\mathcal{F}\left[ \Phi(x,y) \right] \right],
\end{align}
where $\chi(q)$ is the static zero-temperature Lindhard function of a homogeneous noninteracting massless Dirac fermion fluid, $\epsilon(q)$ is the static random-phase-approximation dielectric function and $\mathcal{F}[\ldots]$ denotes a Fourier transform. We have checked that for our problem $\delta\!n$ is negligible and thus the scalar potential $\Phi$ can be ignored.

In Fig. \ref{fig:pseudomag_field_theta}, pseudo-magnetic field $\B$ seen by electrons in the vicinity of the $K_{+}$ valley ($\xi=1$) for a square membrane of length $L=2\mathrm{\mu}$m and $\frac{dL}{L}=1\%$ is shown in relation to the angle $\theta$ (the pseudo-magnetic field for electrons in the other valley has the opposite sign). As the angle is varied, the distribution of $\B$ changes drastically. In particular, for $\theta=0$ [Fig. \ref{fig:pseudomag_field_theta}(a)], $\B$ is asymmetric about the $\xmem$ axis. It follows that electrons propagating along the $\ymem=0$ line do not experience any perturbation associated with the pseudo-magnetic fields. However, as $\theta$ increases, this $\B=0$ region in the middle of the flake is destroyed. The areas where the magnitude of $\B$ is significant shift from the centres of individual quarters of the flake towards the left and right edges [Fig. \ref{fig:pseudomag_field_theta}(b) and (c) for $\theta=\frac{\pi}{36}$ and $\frac{\pi}{12}$, respectively] and for $\theta=\frac{\pi}{6}$ [Fig. \ref{fig:pseudomag_field_theta}(d)] $\B$ is concentrated along those edges and antisymmetric about the $\ymem$ axis. This configuration may severely impact the transport through the system if electronic states are quantized into Landau levels as a result of $\B$. To inspect the latter possibility, we look at the magnitude of the `magnetic length' $\lambda_{\B}$ associated with the field $\B$, $\lambda_{\B}=\sqrt{\frac{\hbar}{e\B}}$, as an indicator of the stability of the Landau level formation: if the pseudo-magnetic field $\B$ changes little at the distance comparable with local $\lambda_{\B}$, the Landau level may form in that portion of the flake. For that purpose, three points on the flake, marked with black, green and orange dots, have been chosen in Fig. \ref{fig:pseudomag_field_theta} and a circle with a radius corresponding to the magnetic length $\lambda_{\B}$ at those points have been drawn using respective colours, for each of the graphs in Fig. \ref{fig:pseudomag_field_theta}. We see that in the case of $\theta=\frac{\pi}{6}$, quantization of electron states into Landau levels may be indeed expected along the left and right edges of the flake: the orange and black circles in Fig. \ref{fig:pseudomag_field_theta}(d) cover regions of reasonably smooth pseudo-magnetic field $\B\sim 0.1$T. For monolayer graphene, that suggests a gap of the order of 10meV between the zero-energy Landau level and the next quantized state. To compare, in bilayer graphene (within our model, the distribution of $\B$ is the same irrespectively of the number of graphene layers in the flake), the gap would be only of the order of 0.5meV. In contrast, no pronounced quantization of the electronic states should be expected for the flake orientation at $\theta=0$. 

\begin{figure}[t]
\centering
\includegraphics[width=0.5\columnwidth]{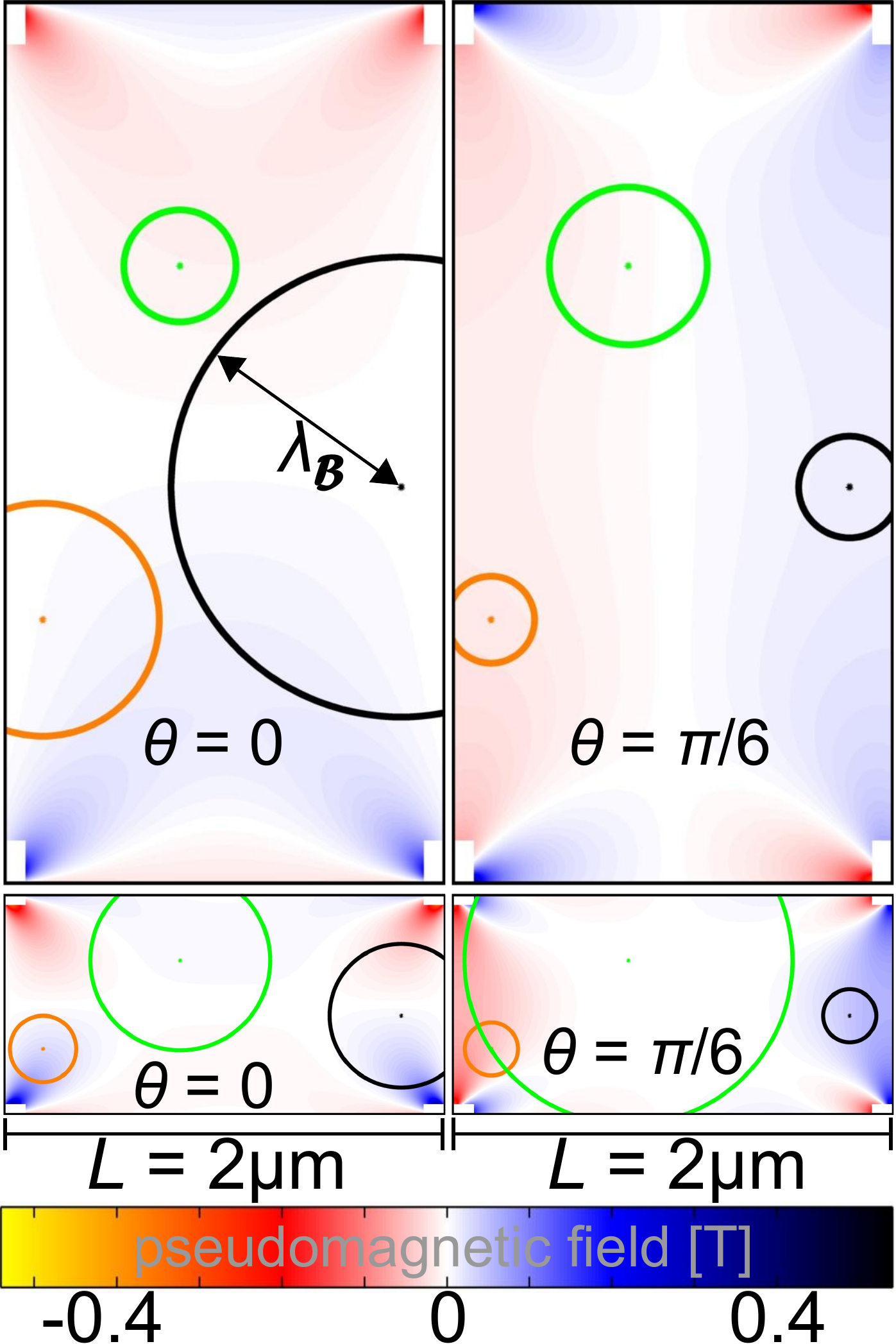}
\caption{Distribution of the pseudo-magnetic field $\B$ for electrons around the $K_{+}$ valley for flakes with aspect ratios $\frac{W}{L}=2$ (top) and $\frac{W}{L}=0.5$ (bottom). Like previously, $\B$ in the vicinity of the corners is ignored. In all cases, $\frac{dL}{L}=1\%$. Distributions for both $\theta=0$ and $\theta=\frac{\pi}{6}$ are shown. The black, green and orange circles show the magnetic length for three points in the membrane marked by a dot in the same colour.}
\label{fig:pseudomag_field_aspectratio}
\end{figure}

We now investigate the dependence of the distribution of the pseudo-magnetic field $\B$ on the aspect ratio of the suspended flake. In Fig. \ref{fig:pseudomag_field_aspectratio}, the strain-induced distribution of $\B$ is shown for flakes with aspect ratios $\frac{W}{L}=2$ and $\frac{W}{L}=0.5$. We consider two most characteristic orientations of the flake with respect to the $\xmem$ axis, the `zig-zag' ($\theta=0$) and `armchair' ($\theta=\frac{\pi}{6}$) orientations. Again, we choose three points on the flake for which a circle of radius equal to the local magnetic length $\lambda_{\B}$ is drawn. For flakes in zig-zag orientation, a possibility of Landau level formation in some areas is indicated in the case of wide and short samples. However, due to symmetry, for such orientation the top and bottom regions of nonzero $\B$ are always separated by a region of $\B=0$. For the armchair orientation, the regions of significant $\B$ are located in the vicinity of the contacts. Irrespectively of the flake orientation, for very wide or very narrow flakes, the centre of the flake is far enough from the corners to perceive the strain as homogeneous and such strain does not generate any pseudo-magnetic field \cite{guinea_natphys_2009, guinea_prb_2010, mucha-kruczynski_prb_2011}.

In summary, we have investigated the distribution of the deformation-induced pseudo-magnetic field in suspended graphene flakes stretched along two opposite edges which remain clamped\footnote{In our analysis, we modeled the flakes as planar membranes and hence neglected the presence of ripples in suspended graphene samples \cite{meyer_nature_2007} as well as any wrinkles forming due to the inhomogeneous stresses within the flake \cite{cerda_nature_2002, cerda_prl_2003}. Out-of-plane deformations enter into the expression for the gauge potential $\vect{\mathcal{A}}_{0}$ via terms like $\left(\partial_{ij}h\right)^{2}$ (Ref. \cite{kim_epl_2008, guinea_prb_2008, guinea_ssc_2009}), where $h(\xmem,\ymem)$ describes the out-of-plane position of a point $(\xmem,\ymem)$ in the membrane. Hence, for dimensions considered in this paper, intrinsic ripples of a graphene flake will only slightly modulate the distribution of the pseudo-magnetic field presented above. Wrinkling is caused by compressive stresses arising within the membrane and develop for strains greater than the critical strain \cite{cerda_nature_2002, cerda_prl_2003, puntel_jelast_2011}. Nonlinear theory neeeds to be applied to fully assess how wrinkles would affect pseudo-magnetic field $\B$ in flakes under tension as discussed here and whether any Landau level quantization may be possible in the regions along the clamped edges, where wrinkles decay.}. We show that details of the distribution of $\B$ depend on the microscopic orientation of the flake with respect to the contacts. In particular, the zig-zag orientation always possesses a $\B=0$ region connecting both contacts in the middle of the flake. For the armchair orientation, similar region connects the free top and bottom edges of the flake and the pseudo-magnetic flake is concentrated in the vicinity of the contacts. We suggest that for the latter, Landau levels may form along the left and right edges due to the presence of $\B$, especially in the flakes with aspect ratios $\frac{W}{L}\lesssim 1$. Such situation would create within the flake regions with gapful electronic spectrum and we estimated the magnitude of this gap as $\sim10$meV for square monolayer graphene flakes with the side length of 2$\mu$m.

We thank A.~Geim and F.~Guinea for useful discussions. This research has been funded by EPSRC S\&IA Grant EP/G035954, EU STREP ConceptGraphene and in part by the National Science Foundation under Grant No. NSF PHY11-25915. M.~M.-K.~acknowledges funding through the Lancaster University Small Grant Scheme. This work has been finalized during the Physics of Graphene program at the KITP, Santa Barbara.


\end{document}